\documentclass[a4paper,12pt]{article}
\usepackage{epsfig}

\title{Overcoming priors anxiety}

\author{Giulio D'Agostini \\
 \mbox{\it Dipartimento di Fisica dell'Universit\`a ``La Sapienza''} \\
 \mbox{\it and Istituto Nazionale di Fisica Nucleare (INFN)}\\
\mbox{\it P.le Aldo Moro 2, I-00185 Roma (Italy)}
}

\date{}

\begin{document}

\maketitle
\begin{abstract}

\footnotetext{Invited contribution to the monoghaphic issue 
of the {\it Revista de la Real Academia de Ciencias} 
on {Bayesian Methods in the Sciences} (J.M. Bernardo ed.). \\
Email: dagostini@roma1.infn.it.\
URL: http://www-zeus.roma1.infn.it/$^\sim$agostini/ }
The choice of priors may become an insoluble problem 
if priors and Bayes' rule are not seen 
and accepted in the framework of subjectivism. 
Therefore, the meaning and the role of subjectivity in 
science is considered and defended from the 
pragmatic point of view
of an ``experienced scientist''. 
The case for the use of subjective priors is then supported
and some recommendations for routine and frontier measurement  
applications are given. The issue of reference priors 
is also considered from the practical point of view 
and in the general 
context of ``Bayesian dogmatism''. 
\end{abstract}

\newpage
\section{Introduction}
The main resistance scientists have  
to Bayesian theory seems to be due to their 
reaction in the face  of  words such as ``subjective'', ``belief''
and ``priors'' (to which the word ``bet'' might also be added). 
These words sound blasphemous to those 
who pursue the ideal of an objective Science. 
Given this premise, it is  not  surprising that
frequentistic ideas, which
advertise objective methods at low cost 
in a kind of demagogical way, became popular very quickly 
and are still the most widely used in all fields of application,
despite the fact that they are indefensible from a 
rational point of view. As in commercials, what often 
matters is just the slogan, not the quality of the product, at least
in the short term. And advertised objective methods are certainly
easier to sell than subjective ones. When one adds 
to these psychological effects
yet others  based upon political 
reasons (see, for example, the very interesting philosophical
and historical introduction to Lad's book\,\cite{Lad}),
life gets really hard for 
subjective probability. 
Moving from the slogan to the product, it is not
difficult to see that, if they were to be taken literally,
 frequentistic ideas would lead nowhere.
Indeed their success seems  due 
to a mismatch between what they state and how
scientists interpret them in good 
faith. In other words, frequentistic 
methods make sense only if they are - when they can be - 
reinterpreted from a subjective point of view. Otherwise 
they may cause serious mistakes to be made. 
In  recent years 
I have investigated  this question among 
particle physicists\,\cite{dagocern,maxent98}. 
For the  convenience of the reader,
I report really  here the main conclusions which 
I  reached\,\cite{maxent98}:
\begin{quote}
{\small\sf 
\begin{itemize}
\item[-]
there is a contradiction between a cultural background in statistics
and the good sense of physicists; physicists' intuition is closer 
to the Bayesian approach than one might na\"\i vely think;
\item[-]
there are cases in which good sense alone is not enough
and serious mistakes can be made; it is then that the philosophical
and practical advantages offered by the Bayesian approach
become of crucial importance;
\item[-]
there is a chance that the Bayesian approach can become 
widely accepted, if it is presented in a way which is close 
to  physicists' intuition and if it can solve the ``existential''
problem of reconciling two aspects which seem irreconcilable:
subjective probability and the honest ideal of objectivity
that scientists have.
\end{itemize}
}
\end{quote}
This last point was just sketched in the original paper, 
and I would like to 
discuss it here in a bit more  detail, 
and to relate it to the ``problem'' of  priors, 
the main subject of this article. 
I think, in fact, that it is impossible to talk about 
priors without putting them into the framework to which they 
belong. Only when one is aware of the role 
they have in Bayes' theorem, and of the role of Bayes' 
theorem itself, can one  have a relaxed relationship with them. 
Once this is achieved, depending on the specific problem, 
one  may choose 
the most suitable priors or  ignore them if they are irrelevant;
or one may decide, instead, that priors are so relevant 
that only Bayes' factors can be provided; alternatively
one may even skip the Bayes 
theorem altogether, or use it in a reverse mode to 
discover which kind of
of priors might give rise to the final beliefs that one 
unconsciously has.  These situations will be illustrated 
by examples. 

Before going any further, 
some clarifications are in order. First, 
my comments will be from the viewpoint of the ``experienced 
scientist'' (i.e. the scientist who is  used
to everyday confrontation with real data);  this  point of view is
often neglected, since priors (and questions of subjectivity/objectivity)
tend to be debated among mathematicians, statisticians and philosophers.
Second, since I am an experimental particle physicist, 
I am aware that my knowledge about the literature
concerning the arguments I am going to talk about  
is necessarily limited and fragmentary. I therefore  apologize if 
people who may have expressed opinions similar to those stated 
in this paper are not acknowledged here.  

\section{Subjective degrees of belief and objective Science} 
The question {\it ``can subjective degrees 
of belief build an objective Science?''}
is subtle.
If we take it literally,
the answer is NO. But this is not because of the 
subjective degrees of belief in themselves.
It is simply because, 
from a logical point of view,
``objective Science'' is a contradiction in terms, if 
``Science'' stands for Knowledge concerning Nature, and ``objective''
for something which has the same logical strength as
a mathematical theorem. 
 This has been pointed out many times by
philosophers, the strongest defence of this point of view
being due to Hume\,\cite{Hume}, to whom there is little to reply.
 
If, instead,  ``objective Science'' stands for what scientists
refer to by this expression, the question becomes a tautology. 
In fact, using Galison' words\,\cite{Galison}, 
{\it ``experiments begin and end in a matrix of beliefs. \ldots
beliefs in instrument types, in programs of experimental enquiry,
in the trained, individual judgements about every local behaviour
of pieces of apparatus\ldots''}.
Any scientist knows already that the only objective 
thing in science is the reading of digital scales. When 
we want to transform this information into scientific 
knowledge we have to make use of many implicit and explicit
beliefs.

 However, many scientists are reluctant
to use the word ``belief''\footnote{But many other 
scientists, usually 
prominent ones, do. And, paradoxically, 
objective science is, for those who avoid the word "belief", 
nothing but the set of beliefs held by the
most influential scientists in whom they believe\ldots} 
for professional purposes.  
It seems to me that the reason for this attitude 
is due to a misuse of the word ``belief'',
which has somehow led to a deterioration of its meaning. In 
this connection I think a few remarks are of 
particular importance. The first is that 
we should have Hume's 
 distinction between ``belief'' and ``imagination'' clear in 
mind\,\cite{Hume}. Then, once we agree on what ``belief'' is,
and on the fact that it can have a degree, and that this degree
depends necessarily on the subject who evaluates it, another 
important concept which enters the game is 
that of de Finetti's ``coherent bet''\,\cite{definetti}. 
The ``coherent bet''  plays the crucial role of
neatly separating  ``subjective'' from ``arbitrary''. In fact,  
coherence has the normative role of   
 forcing people to be honest and to make the best 
(i.e. the ``most objective'') assessments of their degree of 
belief\footnote{The coherence is also important to avoid
making the confusion between ``belief'' and ``convenience'' 
(or ``wish''). In other words, the tasks of assessing 
probability and of decision making should be kept separate.}.
Finally comes Bayes' rule\,\cite{Bayes}, which is the
logical tool  for updating degrees of belief.

In my opinion there is a really good chance that 
this way of presenting the Bayesian theory 
will be accepted by scientists.
In fact the ideal of objectivity is easily recovered, 
although in terms of {\it intersubjectivity}, 
if scientific  knowledge is 
regarded as a very 
{\it solid Bayesian network}\,\cite{Pearl} 
(Galison's 
``matrix of beliefs''\,\cite{Galison}),
based on centuries of experimentation, with {\it fuzzy borders}
which correspond to the areas of current investigation. 

\section{Choosing priors: fear, 
misconception and good faith} 
Once we have specified the exact meaning of each of 
the ingredients  entering probabilistic induction
(degree of belief - coherent bet - Bayes' rule), 
there should, in principle,
no longer be a problem. However, all Bayesians know by experience
that the most serious concerns  scientists have 
are related to the choice of priors
(sometimes due to real technical problems, but
more often due only to ``prejudices on priors'').
In fact, practitioners can avoid talking about ``degree of belief'' 
in their papers, 
replacing it by the nobler term ``probability''; they can
accept the  use of Bayes' theorem, because it is a theorem; but it seems
they cannot escape from priors. And they often get stuck, or simply
go back to ``objective'' frequentistic methods.
In fact, the choice of the prior 
is usually felt to be a vital
problem by all those who approach the Bayesian methods 
with a purely utilitarian spirit, that is, without having assimilated 
the spirit of subjective probability. Some 
use ``Bayesian formulae'' simply because they
``have been proved'', by Monte Carlo simulation, 
to work in a particular application. Others seem convinced
by the power of Bayesian reasoning, 
but they are 
embarrassed because of the apparent ``arbitrariness'' 
of the choice of priors. 

It might seem that  
reference priors (see e.g. \cite{BS} and references therein, although 
in this paper I will refer only to Jeffreys' priors\,\cite{Jeffreys},
the most common in Physics applications) 
have a chance of attracting people to  
Bayesian theory. In fact, reference priors enable  
practitioners  to avoid 
the responsibility for choosing priors, and give them   
 an {\it illusion of objectivity}
analogous to that offered by frequentistic procedures\,\cite{BB}. 
However I have some perplexity
about uncritical use of reference priors, for philosophical, 
sociological and practical reasons which I am now going to explain. 

\subsection{Bayesian dogmatism and its dangers}
Although I agree, in principle,  
that a
{\it  ``concept of a `minimal informative' prior specification
- appropriately defined!''}\,\cite{BS} is valid, 
those who are not fully aware 
of the intentions and limits of  reference analysis perceive
the Bayesian approach  to be dogmatic. 
Indeed, one can find
indiscriminate use and uncritical recommendation of 
reference priors in 
books, lecture notes, articles and conference proceedings 
on Bayesian theory and applications. This gives to practitioners
the impression that only those priors blessed by the official 
Bayesian literature
are valid. This would be a minor problem if the use of reference
priors, instead of more motivated ones, merely caused a 
greater or lesser difference in the numerical result. However,
the question becomes more serious when the - perhaps unwanted - dogmatism
is turned against the Bayesian theory itself. 
I would  like to 
give  an example of this kind which concerns me very much, 
because it may influence the 
High Energy Physics community 
to which I belong. In a paper which appeared last year in   
{\it Physical Review}\,\cite{CF} 
 it is stated that 
\begin{quote}
{\small \sf ``For a parameter  $\mu$ which is restricted to $[0,\infty]$, 
a common non-informative prior in the statistical literature is
$P(\mu_t)=1/\mu_t$\ldots In contrast the 
PDG}\footnote{PDG stands for ``Particle Data Group'', 
a committee that  every second year publishes the {\it Review 
of Particle Properties}\,\cite{PDG}, a very influential collection
of data, formulae and methods, including sections on 
Probability and Statistics.}{\small \sf
description is equivalent to using a prior which is uniform 
in $\mu_t$. This prior has no basis that we know of in Bayesian
theory.''}
\end{quote}
This example should be taken really very seriously. The authors, in fact, 
use the pulpit of a prestigious 
journal to make it seem 
as if they  understand deeply
 both the Bayesian approach and the frequentistic 
approach and, on this basis, they discourage the use of Bayesian methods 
({\it ``We then obtain confidence intervals which are never 
unphysical or empty. Thus they remove an original intention
for the description of Bayesian intervals 
by the PDG''}\,\cite{CF}).

So
it seems 
to me that there is a risk that indiscriminate use of 
reference priors might harm
the Bayesian theory in the long term,
in a similar way to that which happened at the end of 
last century, as a consequence of the abuse of  the uniform distribution. 
This worry is well expressed in Earman's  conclusions 
to his 
``critical examination
of Bayesian confirmation theory''\,\cite{Earman}:
\begin{quote}
{\small\sf ``We then seem to be faced with a dilemma. 
On the one hand, Bayesian considerations seem 
indispensable in formulating and evaluating scientific inference. 
But on the other hand, the use of the full
Bayesian apparatus seems to commit the user to a form of 
dogmatism''.}
\end{quote}
\subsection{Unstated motivations behind Jeffreys' priors?}
Coming now to the specific case of Jeffreys' priors\,\cite{Jeffreys}, 
I must admit that, 
from the most general (and abstract) point of view, it 
is not difficult 
to agree that {\it ``in one-dimensional continuous regular problems,
Jeffreys' prior is appropriate''}\,\cite{BS}. Unfortunately, it is
rarely
the case that in practical situations
the status of prior knowledge 
is equivalent to that expressed by the Jeffreys' priors, as I will 
discuss later. Reading ``between the lines'', 
it seems to me that the reasons for choosing these priors are essentially 
psychological and sociological. For instance, 
when utilized to infer $\mu$ (typically 
associated with the ``true value'') from
``Gaussian small samples'', the use of a prior of the kind 
$f_\circ(\mu,\sigma)\propto 1/\sigma$ has two apparent benefits:
\begin{itemize}
\item
first, the mathematical solution is simple (this reminds 
me of the story of the drunk under the streetlamp, 
looking for the key lost in the dark alley);
\item 
second, one recovers the Student distribution, and for some 
it seems to be reassuring that a Bayesian result gets blessed by 
{\it ``well established''} frequentistic methods. 
(``We know that this is the right solution'', 
a convinced Bayesian once told me\ldots)
\end{itemize}
But these 
arguments, never explicitly stated, cannot be accepted, 
for obvious reasons. I would like only to comment on the 
Student distribution. This is 
the ``standard way'' for 
handling small samples, although  
 there is, in fact, no deep reason 
for aiming to get such a distribution for the posterior. 
This becomes clear to 
anyone who, having measured  the size of 
this page twice and having found a difference 
of 0.3 mm between the measurements, then has to base  
his conclusion on that distribution. 
Any
rational person  
will refuse to state that, in order 
to be 99.9\,\% confident in the result,
the uncertainty interval should be 9.5 cm wide
(any carpenter would laugh\ldots). This might be the
reason why, as far as I know, physicists don't
use the Student distribution.  

Another typical application of the Jeffrey' prior is in the 
case of inference on the $\lambda$ parameter 
of a Poisson distribution, having observed a 
certain number of events $x$.
Many people have, in fact, a reluctance
to accept, as an estimate of $\lambda$, a value which differs from 
the observed number of counts 
(for example, $\mbox{E}(\lambda)=x+1$ starting from a uniform
prior) and which is deemed to be distorted 
by the ``distorted''
 frequentistic criteria used to analyse the problem
 (see e.g. \cite{dagocern}). 
In my opinion, in this case one should simply
 educate the practitioners about the difference
between the concept of maximum belief and that of {\it prevision}
(or expected value).  An example in which the choice of
priors becomes
 crucial, is the case where no counts are observed, 
a typical situation for frontier physics, 
where new and rare 
phenomena are constantly looked for. Any reasonable prior
consistent with what I like to call the ``positive
attitude of the physicists who have pursued the 
research''\,\cite{dagocern}, 
allows 
reasonable upper limits  compatible with the sensitivity 
of the experiment to be calculated (even a uniform prior
is good for the purpose).
Instead, a prior of the kind 
$f_\circ(\lambda)\propto1/\lambda$
prevents  use  of  probabilistic statements to
summarize the outcome  of the experiment, 
and the same result ($0\pm0$) is obtained, 
independently of the size sensitivity and running
time of the experiment.

I will return below to such critical situations which are 
typical of frontier science. 

\section{Priors for routine applications}
Let us discuss now the reasons 
which indicate that experimentally motivated priors for 
``routine measurements''
are quite different from Jeffreys' priors. 
 This requires a  brief reminder about 
how measurements are actually performed. 
I will also take the opportunity
to introduce the  International Organization for 
Standardization (ISO) recommendations concerning measurement 
uncertainty. 
\subsection{Unavoidable prior knowledge behind any measurement}
To understand why an ``experienced scientist''
 has difficulty in
 accepting 
a prior of the kind $f_\circ(\sigma) \propto 1/\sigma$
(or $f_\circ(\ln(\sigma)) = k$), one has to remember
 that the process 
of measurement is very complex (even in everyday situations, 
like measuring the size of the page 
\underline{You} are reading 
\underline{now}, 
just to avoid abstract problems): 
\begin{itemize}
\item
 first one has to {\it define the measurand}, i.e. the quantity 
 one is interested in;
\item
 then one has to {\it choose the appropriate instrument}, 
one which has
known properties,  well-suited range and resolution, 
and in which one has some confidence, 
achieved on the basis of previous
measurements;
\item 
 the {\it measurement} is performed and, if possible, 
repeated several times;
\item 
then, if one judges that this is appropriate, 
one applies  {\it corrections}, 
also based on previous experience with that kind of
measurement, in order to take into account
known (within uncertainty) systematic errors; 
\item 
finally\footnote{This is not really the end of the story, 
if a researcher wishes his  
result to have some impact on the scientific community.
Only if other people trust him will they
use the result in further scientific 
reasoning, as if it were their own result. This is the reason 
why one has to undergo an apprenticeship 
during one's youth, when one must 
build up one's reputation
(i.e. again beliefs) in the eyes of
one's colleagues.} 
one gets a credibility interval for the quantity
(usually a {\it best estimate} with a related {\it uncertainty});
\end{itemize}
Each step involves some prior knowledge and, typically,
each person who performs the measurement 
(be it a physicist, a biologist, or a carpenter)
operates in his field of expertise. This means that he
is   well aware of the 
error he might make, and therefore of the uncertainty
associated with the result. This is also true if only 
a single observation has been 
performed\footnote{This defence of the possibility 
of quoting an uncertainty from a single measurement 
has nothing to do with 
the mathematical games like those of \cite{Rodriguez}.}:
try to ask 
a carpenter how much he believes in his result, 
possibly helping him to quantify the uncertainty 
using the concept of the coherent bet. 

There is also another important aspect of the 
``single measurement''. One should  
note that many measurements, which seem to be due to a single observation, 
consist, in fact, of several observations 
made within a short time: for example, 
measuring a length with a design ruler, one checks 
the alignment of the zero mark
 with the beginning 
of the segment to be measured several times;
or, measuring a voltage with a voltmeter or 
a mass with a balance, one waits until the reading 
is well stabilized. Experts  use unconsciously also 
information of this kind when they have to 
figure out the uncertainty  they attribute to the result,
although they are unable to use it explicitly because 
this information cannot be accommodated 
in the standard way of evaluating uncertainty  
based on frequentistic methods\,\cite{maxent98}.

The fact that the evaluation of uncertainty does  not necessarily  
come from repeated measurements has also been 
recognized by the International Organization for Standardization
(ISO)
in its {\it ``Guide to the expression of uncertainty in 
measurement''}\,\cite{ISO}. 
There the uncertainty
is classified 
\begin{quote}
{\small\sf ``into two categories according to the way their 
numerical value is estimated:
\begin{enumerate}
\item[A.] those which are evaluated by 
          statistical methods\footnote{Here ``statistical'' 
should be seen as referring to 
``repeated observations on the same measurand'', and not to
general meaning of ``probabilistic''.};
\item[B.] those which are evaluated by other means;''  
\end{enumerate}
}
\end{quote}
Then, illustrating the ways to evaluate the  
``type B standard uncertainty'', the {\it Guide} states that
\begin{quote}
{\small \sf ``the associated estimated variance $u^2(x_i)$ or the standard
uncertainty $u(x_i)$ is evaluated by scientific judgement based on all
of the available information on the possible variability of $X_i$. 
The pool of information may include
\begin{itemize}
\item[-] previous measurement data; 
\item[-] experience with or general knowledge of the behaviour
and properties of relevant materials and instruments;
\item[-] manufacturer's specifications;
\item[-] data provided in calibration and other certificates;
\item[-] uncertainties assigned to reference data taken from handbooks.''
\end{itemize}
}
\end{quote}
It is easy to see that the above statements have sense
only if the probability is interpreted as degree of belief,
as explicitly recognized by the {\it Guide}:
\begin{quote}
{\small \sf
``\ldots Type B standard uncertainty is obtained from an 
assumed probability density function based on the degree of belief that 
an event will occur [often called subjective probability\ldots].'' 
}
\end{quote}
It is also interesting to read the
concern of the {\it Guide} regarding the uncritical use of 
statistical methods and of abstract formulae:
\begin{quote}
{\small\sf ``the evaluation 
of uncertainty is neither a routine task nor a 
purely mathematical one; it depends on detailed knowledge
of the nature of the measurand and of the measurement.
The quality and utility of the uncertainty quoted for
the result of a measurement therefore ultimately 
depend on the understanding, critical analysis, 
and integrity of those who contribute to the assignment 
of its value''.
}
\end{quote}
This appears to me perfectly in line with the lesson
of  genuine subjectivism, accompanied by the 
normative rule of  coherence. 

\subsection{Rough modelling of realistic priors}
After these comments on measurement, it should be clear why 
a prior of the kind $f_\circ(\mu,\sigma)\propto 1/\sigma$
does not look natural.  
As far as $\sigma$ is concerned, this prior would imply, in fact,  that 
standard deviations ranging over  many (``infinite'', in principle)
 orders of magnitude 
would be equally possible. This is unreasonable in most cases. 
For  example, measuring the size of this
page with a design ruler, no one 
would expect  $\sigma\approx {\cal O}(10\,\mbox{cm})$ or
$\approx {\cal O}(1\,\mu\mbox{m})$. As for $\mu$, the choice
$f_\circ(\mu)=k$ is acceptable until $\sigma\ll \mu$ (the 
so called Savage {\it principle of precise 
measurement}\,\cite{Savage}). But when the order of magnitude 
of $\sigma$ is uncertain, the prior on $\mu$  
should also be revised (for example, most of the directly measured
quantities are positively defined).  

Some priors which, in my experience, are closer 
to the typical prior knowledge of the person 
who makes {\it routine measurements} 
are those concerning the order of magnitude of $\sigma$, or the order
of magnitude of the precision (quantified by the 
variation coefficient $v=\sigma/|\mu|$).  
For example\footnote{For the sake of simplicity, let us stick to the case 
in which the fluctuations are larger than the intrinsic 
instrumental resolution. Otherwise one needs to model 
the prior (and the likelihood) with a discrete distribution.},
one might expect a r.m.s. error of 1 mm, but 
values of  0.5 or 2.0 mm would not look surprising. Even 
0.2 or 4 mm would look possible, but certainly not
1\,$\mu$m or $10\,$cm. So, depending on whether 
one is uncertain on the absolute or the relative error,
a distribution which seems suitable for a rough modelling 
of this kind of prior is a {\it lognormal} in either $\sigma$ 
or $v$. For instance, the above example could be modeled 
with $\ln{\sigma}$ normally distributed with average 0
 ($=\ln 1$) and standard deviation 0.4. The 1, 2 and 3 standard 
deviation interval on $\sigma$/mm would be 
$[0.7, 1.5]$, $[0.5, 2.2]$ and $[0.3, 3.3]$, respectively, 
in qualitative agreement with the prior knowledge. 

In the case of more sophisticated measurements, in which the 
measurand is a positive defined quantity 
 of unknown order of magnitude, a suitable prior would   
again be a normal (or, at limit, a constant)
 in $\ln \mu$ (before the first measurement one may be uncertain 
on the order of magnitude that will be obtained), while 
$\sigma$ is somehow correlated to $\mu$ 
(again $v$ can be reasonably described by a lognormal). 
One might imagine of other possible measurements which give rise 
to other priors, but I find it very difficult to imagine 
a real situation for which Jeffrey's priors are 
appropriate. 
\subsection{Mathematics versus good sense}
The case of small samples seems to lead to an impasse. 
Either we have a simple and standard solution
to a fictitious problem,
given by the 
Student distribution,  
or we have to face complicated calculations 
if we want to solve specifically the problem we have in mind, 
formulated by modeling experience motivated priors.
I do not think that experimenters would be  willing to  
calculate lognormal integrals to report the results of a couple 
of measurements. This could be done once, perhaps, to get a feeling 
of what is going on, or to solve an academic exercise, 
but certainly not as routine. 

The suggestions sketched above were in the 
framework of the Bayes' theorem
paradigm. But I don't want to give the impression
that this is the only way to proceed.
The most important teaching of subjective probability 
is that probability is always conditioned by a given status 
of information. The probability is updated in the light 
of  any new information. But it is not always possible to 
describe the updating mechanism using the neat scheme of 
the Bayes' theorem. This is well known in many fields, 
and, in principle, there is no reason  for considering the 
use of the Bayes theorem to be indispensable to 
assessing uncertainty in scientific measurements. The idea is to force the 
expert to declare (using the coherent bet) some quantiles 
in which he believes is contained the true value, on the basis
of a few observations. It may be easier for
him to estimate the uncertainty in this way, drawing on his 
past experience, rather than trying to model some priors and 
playing with 
the Bayes' theorem. The message is what experimentalists
intuitively do: {\it when you have just a few observations,
what you already know is more important than what 
the standard deviation of the data teaches you}.

Some will probably be worried by the arbitrariness of this conclusion,
but it has to be remembered that: an expert 
can make very good guesses in his field;  20, 30, or even 50 \%
uncertainty in the uncertainty is not considered as
significantly spoiling the quality of a measurement; there 
are usually many other sources of uncertainty, due to 
possible systematic effects of unknown size, which 
can easily be more critical. 
I am much more worried by the attitude of giving up
prior knowledge to a mathematical convenience, since this 
can sometimes lead to paradoxical results. 

\subsection{Uniform prior for $\mu$, $\lambda$ and $p$ in routine 
measurements}
I find, on the other hand, that for routine applications
the use of the uniform distribution for the center parameter
of the normal distribution, usually associated to the true
value, is very much justified.
This is because, apart from pathological situations, 
or from particular cases in frontier research, even if one does
not know if the associated uncertainty will be 0.1, 1, or 10\,\%, 
the prior knowledge on $\mu$ is so vague that it can be 
considered uniform for all practical purposes. The same holds 
when one is interested in $\lambda$ of a Poisson distribution
(counting experiments) or to $p$ of the binomial distribution
(measurements of proportions), under the condition that 
\underline{normal approximation} is roughly satisfied, which is a 
kind of desideratum for the planning of a good routine experiment
(otherwise it becomes a non-routine one).  
Taking into account the fact that, for routine 
measurements,
the difference between mode and
average of the final distribution is much smaller than 
$\sigma(\lambda)$ or $\sigma(p)$, we ``recover'' maximum likelihood
results, but with a natural, i.e. subjective, 
interpretation of the results. 
This corresponds, in fact, to the case where the intuitive 
``dog-hunter probability inversion''\,\cite{dagocern,maxent98} is reasonable.
For example, indicating by $x$ the number of observed events in the case
of Poisson distribution or of successes in the case of the binomial one,
with the number of trials of the latter indicated by $n$, we get, simply
\begin{eqnarray}
\lambda &\sim & {\cal N}(x, \sqrt{x}) \label{eq:lambda}\\
p       &\sim &  {\cal N}\left(\frac{x}{n},\sqrt{\frac{x}{n}\,
                 \left(1-\frac{x}{n}\right)\,\frac{1}{n}}\right)\label{eq:p}\,,
\end{eqnarray}
where ${\cal N}(\cdot, \cdot)$ is  short hand for 
normal distribution of given average and standard deviation. 

The recommendation I usually like to give, to check ``a posteriori'' 
if the uniform 
prior is suitable or not is the following: first evaluate 
central value and standard deviation according to the approximations 
 (\ref{eq:lambda}) and (\ref{eq:p}); then try to judge if 
the central value ``disturbs'' you, and/or the 
standard deviation seems to be of the order of your prior
vagueness; if this is the case, it is now that you need to 
model down some priors, which will actually affect the posteriors; 
otherwise, priors will have no appreciable 
effect and the approximated result is good enough. 

This ``a posteriori' consideration of priors   might 
seem questionable, but I find it absolutely consistent with the spirit 
of subjective probability. In fact, the priors one has  
plug into  Bayes' theorem should reflect the status of 
knowledge as it is felt to be by the subject who performs the 
inference. But sometimes it can be difficult to model this information
consciously, or it might simply take too much time. 
The comparison of the approximate result got
from a  uniform prior with the result that the researcher 
was ready to accept can help, indeed, to 
raise this status of prior knowledge 
from the unconscious to the conscious. 

\section{Priors for frontier science} 
The question is completely reversed when one is interested 
in quantities whose value  might be at the edge 
or beyond the 
 sensitivity of the experiment
(perhaps even orders of magnitudes beyond it) and  
if the quantity itself makes sense at all. 
This is a typical situation in particle physics or in astrophysics,
and it is only to these kind of measurements that I 
  will refer to as ``frontier science measurements''. 
However, even though they are ``frontier'',  
 most of the measurements performed in the above mentioned
fields belong, in fact, to the class of  ``routine measurements''.
 
\begin{figure}
\begin{center}
\epsfig{file=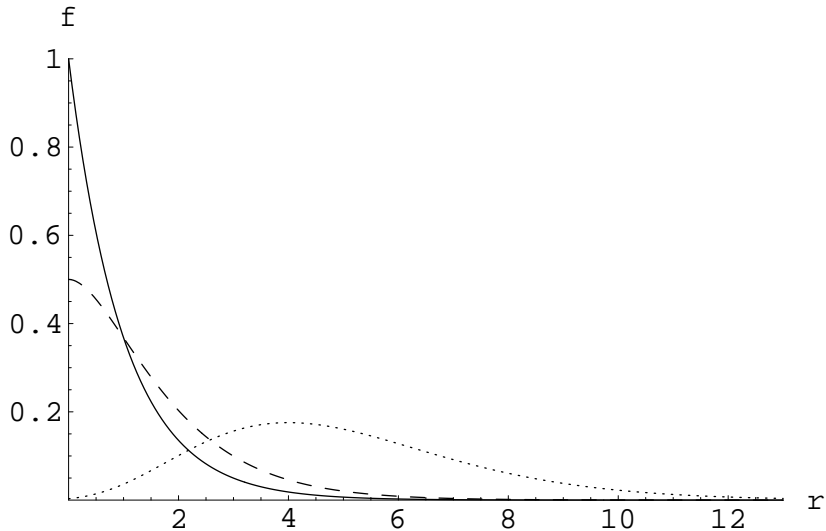,clip=,width=0.80\linewidth} 
\end{center}
\caption{\small Final distribution for the Poisson intensity parameter 
$r$, obtained from a uniform prior and with the following 
values of expected background and observed events: 
0, 0 (continuous); 1, 1 (dashed); 1, 5 (dotted).}
 \label{fig:bayesflat}
\end{figure}

I would like to illustrate this new situation 
with a numerical example. 
Let us imagine that an experiment has been run for one year 
looking for rare events, like magnetic monopoles,  
proton decays, or  gravitational waves. The physics quantity of 
interest (i.e. a decay rate, or a flux) 
is related to the intensity $r$ of a Poisson process. 
Usually there is also an additional Poisson process to be considered, 
associated with the  physical or instrumental background which 
produces observables indistinguishable from the process of 
interest ($r_B$). 
The easy, although ideal, case 
is when the background is exactly zero 
and at least one event is observed. 
This case prompts researchers to make a discovery claim.  
Let us consider, instead,  the situation when no 
candidate events are observed, still 
with zero background. The likelihood, considering 1 year as unit time, 
is $f(x=0\,|\,r)=e^{-r}$. Considering a uniform prior for $r$, 
we get $f(r\,|\,x=0)=e^{-r}$
(see figure \ref{fig:bayesflat}), from which  
a 95\,\% probability {\it upper limit} ($r_u$) can be evaluated. 
This comes out to be $r_u = 3$ events/year and 
it is a kind of standard way in HEP 
of reporting a negative search 
result\footnote{It is worth noting that many physicists are convinced
that the reason for this value is due to the fact that the probability 
of getting 0 from a Poisson of $\lambda=3$ is 5\,\%. This
is  the classical arbitrary probability inversion\,\cite{maxent98} 
which in this case comes out to be correct, assuming a flat prior, 
due to the property of the exponential under integration.}.  
The usual interpretation of this result is that, \underline{if}
the process looked for exists at all, then there is 95\,\% 
probability that $r$ is not greater than $r_u$. But, I find that 
often one does not pay enough attention to all the logical implications
contained in this statement, or in all the infinite probabilistic 
statements which can be derived from $f(r\,|\,x=0)$.
This can be highlighted 
considering statements complementary to the standard ones, especially 
in those cases in which the experimenters feel that the detector 
sensitivity is not suitable for searching for such a rare process. 
The embarrassing reply to questions like {\it ``do you really
believe 5\,\% that $r$ is \underline{greater} than $r_u$?''}, or 
{\it ``would you  really place a 1 to 19 bet on $r > r_u$''}
shows that, often, $f(r\,|\,x=0)$ does not describe coherent beliefs. 
And this is due to the fact that the priors were not appropriate 
to the problem. For example, a researcher  
could run a cheap monopole experiment 
for one day, using a 1 m$^2$ detector, find no candidates and 
present, without hesitation,  
his 95\,\% upper limit as $r_u = 3$\,monopoles/m$^2$/day, 
or 1095\,monopoles/m$^2$/year.  But he would react immediately 
if we made him aware 
that he is also saying that there is 5\,\% chance that the 
monopole flux is above 1095\,monopoles/m$^2$/year, because he 
\underline{knows} that ${\cal O}(1000)$\,m$^2$ detectors 
have been run for many years without observing a convincing signal. 

The situation becomes even more complicated when one  has a non
zero expected  background and a  
number of observed candidates superior to it. For example,
researchers could expect a background of 1 event per day and 
observe 5 events. Differently from the above example of the 
monopole search, let us imagine that the prior knowledge 
is not so strong that all the 5 events can be attributed
with near certainty to background. 
Instead, let us imagine that the experimenters 
are  here in serious trouble: the $p$-value is below 0.5\,\%;   
they  do not believe strongly that the excess is due to 
the searched for effect; but neither do they 
 {\it feel} that the probability is
so  low that they can decide not to publish the result and miss
the chance of a discovery. If they perform a standard 
Bayesian analysis using a flat prior they will get a final distribution 
 peaked at 4 which looks like a convincing signal, since 
it seems to be well separated from 0  
(see figure \ref{fig:bayesflat}). 
They could use, instead, a Jeffreys' prior and find no result, 
since $P(r\le r_\circ)/P(r > r_\circ) =\infty$ for any 
$r_\circ > 0$. It is easy to see that in such a situation 
pedantic use of the Bayesian theory (``Prior, Likelihood 
$\rightarrow$  Final'') leads to an embarrassing 
outcome whatever one does. 

Therefore, in the case of real frontier science observables, 
the best solution seems to be that one has to abstain 
from providing final distributions and publish only likelihoods, 
which are degrees of beliefs too, but they are much less critical 
than priors. But reporting the likelihoods as such can be 
inconvenient, because often
they  do not give an intuitive and direct idea of  the 
power of different experiments.  
Recently, faced with problems of the kind described above, 
I have realized that a  very convenient quantity to use is 
a function that gives the Bayes factor of a generic 
value of interest
with respect to the asymptotic value for which the experimental 
sensitivity is lost (if the asymptotic value exists and the 
Bayes factor is finite)\cite{zeus_ci,higgs,pia}. In the simple
case of the Poisson process with background that we 
are considering, we have
\begin{equation}
{\cal R}(r) \equiv 
\frac{f(x\,|\,r, r_B)}{f(x\,|\,r=0, r_B)} 
\label{eq:r}
\end{equation}
 The advantage 
of this function is that it has a simple intuitive 
interpretation of 
{\it shape distortion function}  of the p.d.f. 
(or a {\it relative belief updating ratio}\footnote{In fact, 
in the case $f_\circ(r)\ne 0$ one can rewrite (\ref{eq:r})
in the following way 
$$
{\cal R}(r)  =
\frac{f(r\,|\,x, r_B)/f_\circ(r)}{f(r=0\,|\,x, r_B)/f_\circ(r=0)}\,.
$$
}) 
introduced by the new observations. 
As long as ${\cal R}$ is 1 it means that the experiment is 
not sensitive and  
the shape of the p.d.f. (and hence the
relative beliefs) remain unchanged. Instead, when ${\cal R}$
goes to zero the beliefs go to zero too, no matter 
how strong they were before. 
Moreover, since the ${\cal R}$ differs from the likelihood 
only by a multiplicative factor, it can be used directly 
in the Bayes' formula when a scientist 
wants to turn them into probabilities, using subjective priors. 
Different experiments can easily be compared and the 
combination of 
independent data is performed multiplying the different ${\cal R}$'s. 

\begin{figure}
\begin{center}
\epsfig{file=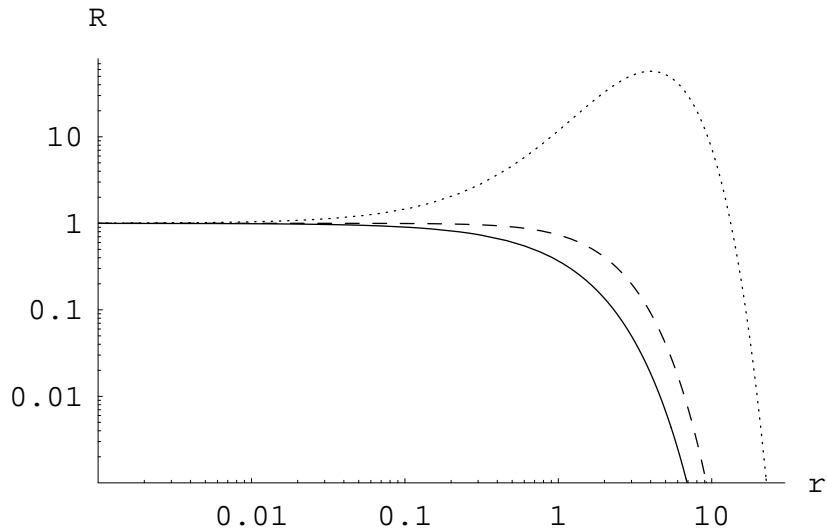,clip=,width=0.80\linewidth} 
\end{center}
\caption{\small Bayes factor  with respect to $r=0$ 
for the Poisson intensity parameter $r$ 
obtained from  the following 
values of expected background and observed events: 
0, 0 (continuous); 1, 1 (dashed); 1, 5 (dotted).}
 \label{fig:R}
\end{figure}

The function ${\cal R}$ is particularly 
intuitive when plotted with the abscissa in log scale. 
For example, figure  \ref{fig:R}, shows the result in terms 
of the ${\cal R}$ function for the same cases shown in figure 
\ref{fig:bayesflat}.
Looking at the plot, one can  immediately get an idea 
of what is going on.
For example, it  also becomes clear where the problems with the 
flat prior and with the Jeffreys' prior come from.    
We can also now understand which kind of priors
the hesitant researchers of the above example had in mind. 
Their prior beliefs were concentrated some order of magnitudes
below the peak of {\cal R},  but with tails which could 
also accommodate
$r\sim {\cal O}(4)$. This is in agreement with the fact 
that after the observations the intuitive probability for 
$r > {\cal O}(1)$  becomes sizable (5, 10, 30\,\%?) and the
researchers do
not have the courage not to publish the result. 

Finally, let us comment on upper (or lower)
limits. It is clear now that,  
exactly in those frontier situations in which 
the limit would be pertinent, a highly intersubjective 
95\,\% probability limit does not exist. Therefore one has 
to be very careful in providing such a quantity. 
However, looking at the plots of figure \ref{fig:R}
it is also clear that one can talk somehow about a bound
of values which roughly separates  the region of ``possible'' values
from that of ``impossible'' ones. One could then take a
conventional value, which could be the value 
for which ${\cal R}=0.05$, or 0.5, or any other. The important
thing is to 
avoid calling this conventional value a 95\,\% or a 50\,\% 
probability limit. 
If  instead one really wants to give a probability limits, 
one has to go through priors, which should be 
precisely stated. In this case, if I really had to recommend 
a prior, it would be the uniform distribution.
This is not, only, for mathematical convenience,
but also because it seems to me that it can do a good
job in many cases of interest. In fact, one can see
that it gives the same result as any other reasonable prior
consistent with the positive attitude of the researchers 
which have planned and financed the experiment\,\cite{dagocern}
(for example, if an experimental
team performs a dedicated proton decay experiment with the intention
of making a good investment of public money, it means that 
the physicists really do hope to observe a reasonable 
amount of signal events for the planned sensitivity of the 
experiment). 

\section{Conclusions}
The key point expressed in this paper 
is that there is no need to ``objectivise''
Bayesian theory, treating subjectivism
as if it were  
something we should  be ashamed of. 
Only when this point is accepted and Bayes' 
theorem is correctly placed
 within the framework of subjective probability, 
with clear role and limitations, can the anxiety
 about priors and their choice  be overcome. 
Once this is achieved, either we can choose 
the priors which best describe  the prior knowledge
for a specific problem; or we can ``ignore'' them in 
routine applications, thus ``recovering'' maximum likelihood results,
but with transparent subjective interpretation, and with
awareness of the assumptions we are using; 
or we can decide that priors are so critical that only 
likelihoods or Bayes factors can be provided as the outcome of
the experiment; 
or we can use the Bayes theorem in a reverse mode,
to find out which  priors we had, unconsciously, 
that give rise to the beliefs we have after the new observation; 
finally there are some cases in which it is even more practical 
to skip the Bayes' theorem and to assess directly 
the degree of belief. With respect to this last point, 
I would like to remind the reader that, in fact,  if one thinks that 
probabilities must only be calculated  using the Bayes' rule,
 one gets trapped in an endless prior-final chain.  

As far as reference priors are concerned, they could, indeed, 
simplify the life of the practitioners in some well 
defined cases. However, 
 their uncritical use should be discouraged. First,  
because they could lead to wrong, or even absurd, results 
in critical situations, if reference priors 
are preferred to case motivated priors just
 for formal convenience. 
Second, and more important, because of they might give
the impression of dogmatism, which, together
with the absurd results obtained through their misuse, 
 could seriously damage 
 the credibility of the Bayesian theory itself.


\begin{thebibliography}{ref99}

\bibitem{Lad}
F. Lad, {\it ``Operational subjective statistical methods - a mathematical,
philosophical, and historical introduction''},
John Wiley \& Sons Ltd,  1996.

\bibitem{dagocern}
G. D'Agostini, {\it ``Bayesian reasoning in High Energy Physics - principles
and applications''}, 
lecture notes
given at CERN (Geneva), May 25--29 1998. CERN Yellow Report
in preparation, to appear at
{\tt http://wwwas.cern.ch/library/cern\_publications/yellow\_reports.html}.
Preliminary version available at the 
author's URL.

\bibitem{maxent98} 
G. D'Agostini, {\it ``Bayesian reasoning versus conventional statistics
in High Energy Physics''},
Proc. of the
XVIII International Workshop on Maximum Entropy and Bayesian
Methods, Garching (Germany), July 1998, V. Dose, W. von der Linden,
R. Fischer, and R. Preuss, eds. (Kluwer Academic Publishers,
Dordrecht, 1999); LANL preprint {\tt physics/9811046}.
A copy can be found 
at the author's URL.

\bibitem{definetti}
B. de Finetti, {\it ``Theory of probability''},
J. Wiley \& Sons, 1974.

\bibitem{Bernardo}
J.M. Bernardo, {\it ``Non-informative priors do not exist''},
J. Stat. Plan. and Inf. {\bf 65}(1997)159, including
discussions by D.R. Cox, A.P. Dawid, J.K. Ghosh and D. Lindley,
pagg. 177-189.

\bibitem{Jeffreys}
H. Jeffreys, {\it ``Theory of probability''}, Oxford
University Press, 1961.

\bibitem{Hume}
D. Hume, {\it ``Enquiry concerning human understanding''} (1748); 
electronic version in  
{\tt http://www.utm.edu/research/hume/wri/1enq/}.

\bibitem{Galison}
P.L.  Galison, {\it ``How experiments end''}, 
The University of Chicago Press, 1987.

\bibitem{Bayes}
T. Bayes, {\it ``An assay toward solving a problem in the
doctrine of chances''} (1764), Philosophical Transactions of the Royal 
Society, 1973, 370-418. For a reproduction see, e.g., 
S. J. Press, 
``{\it Bayesian statistics: principles, models, and applications}'',
John Wiley \& Sons Ltd, 1989. 

\bibitem{Pearl}
J. Pearl, {\it ``Probabilistic reasoning in intelligent systems:
networks of plausible inference''}, Morgan Kaufmann Publishers, 1988.

\bibitem{BS}
J.M. Bernardo, A.F.M. Smith,
``{\it Bayesian theory}'', John Wiley \& Sons Ltd, Chichester, 1994.

\bibitem{BB}
J.O. Berger and D.A. Berry, {\it ``Statistical analysis and the
illusion of objectivity''}, {\it American Scientist} {\bf 76} (1988) 159.

\bibitem{CF}
G.J. Feldman and R.D. Cousins, 
{\it ``Unified approach to the classical statistical 
analysis of small signals''}, {\it Phys. Rev. D} {\bf 57} (1998) 3873.

\bibitem{PDG}
Particle Data Group, R.M. Barnet et al., 
{\it ``Review of particle properties''},
{\it Phys. Rev. D} {\bf 54} (1996) 1.

\bibitem{Earman}
J. Earman, {\it ``Bayes or bust? A critical examination of
Bayesian confirmation theory''}, The MIT Press, 1992. 

\bibitem{Rodriguez}
C.C. Rodriguez, {\it ``Confidence intervals from one observation''},
unpublished 
(paper available in {\tt http://omega.albany.edu:8008/}).

\bibitem{ISO}
International Organization for Standardization (ISO),
{\it ``Guide to the expression of uncertainty in measurement''},
Geneva, Switzerland, 1993.

\bibitem{Savage}
L.J. Savage et al., 
{\it ``The foundations of statistical inference: a discussion''},
Methuen, London, 1962.
 
\bibitem{zeus_ci}
ZEUS Collaboration, J. Breitweg et al., 
{\it ``Search for contact interactions in deep inelastic 
$e^+p\rightarrow e^+ X$ scattering at HERA''}, DESY-99-058,   
  LANL Archive,  
{\tt hep-ex/9905039}, May 1999, submitted to {\it Eur. Phys.  J.} C; 

\bibitem{higgs}
G. D'Agostini and G. Degrassi, {\it ``Constraints on the 
Higgs boson mass from direct searches and precision measurements''}, 
internal report DFPD-99/TH/02; LANL Archive,  
{\tt hep-ph/9902226}, February 1999, 
to be published in {\it Eur. Phys.  J.} C; 
a copy is available at the author's URL. 

\bibitem{pia}
P. Astone and G. D'Agostini,
{\it ``Inferring the intensity of Poisson processes at the 
limit of the detector sensitivity (with a case study 
on gravitational wave burst search)''}, 
paper in preparation.
\end{thebibliography}
\end{document}